\documentclass[
superscriptaddress,
preprint,
prd,tightenlines,showpacs,nofootinbib,
eqsecnum,
amsfonts,amsmath,amssymb]{revtex4-1}

\usepackage{bm}
\usepackage{graphicx}

\newcommand{\nn}{\nonumber}

\newcommand{\ud}{\mathrm{d}}
\newcommand{\uD}{\mathrm{D}}
\newcommand{\ui}{\mathrm{i}}

\newcommand{\calO}{\mathcal{O}}

\newcommand{\ph}[1]{\phantom{#1}}

\newcommand{\cte}{\mathrm{cte}}
\newcommand{\be}{\begin{equation}}
\newcommand{\ee}{\end{equation}}

\begin{document}
\vspace*{4cm} \title{Gravitational waves from spinning compact
  object binaries: \\New post-Newtonian results}

\author{Sylvain MARSAT}

\email{smarsat@umd.edu}
\address{Maryland Center for Fundamental Physics \& Joint Space-Science Institute, Department of Physics, University of Maryland, College Park, MD 20742, USA \\ Gravitational Astrophysics Lab, NASA Goddard Space Flight Center, 8800 Greenbelt Rd., Greenbelt, MD 20771, USA }

\author{Luc BLANCHET}

\email{blanchet@iap.fr}
\address{Institut d'Astrophysique de Paris -- UMR 7095 du CNRS, Universit\'e Pierre \& Marie Curie, 98bis Boulevard Arago,  75014 Paris, France}

\author{Alejandro BOH\'E}

\email{alejandro.bohe@uib.es}
\address{Departament de F\'isica, Universitat de les Illes Balears,\\ Crta. Valldemossa km 7.5, E-07122 Palma, Spain}

\author{Guillaume FAYE}

\email{faye@iap.fr}
\address{Institut d'Astrophysique de Paris -- UMR 7095 du CNRS, Universit\'e Pierre \& Marie Curie, 98bis Boulevard Arago,  75014 Paris, France}


\begin{abstract}
We report on recent results obtained in the
  post-Newtonian framework for the modelling of the gravitational
  waves emitted by binary systems of spinning compact objects (black
  holes and/or neutron stars). These new results are obtained at the
  spin-orbit (linear-in-spin) level and solving Einstein's field equations iteratively
   in harmonic coordinates as well as the
  multipolar post-Newtonian formalism. The dynamics of the binary was
  tackled at the next-to-next-to-leading order, corresponding to the
  3.5 post-Newtonian (PN) order for maximally spinning objects, and
  the result is found to be consistent with a previously obtained
  reduced Hamiltonian in the ADM approach. The corresponding
  contribution to the energy flux emitted by the binary was obtained
  at the 3.5PN order, as well as the next-to-leading 4PN tail
  contribution to this flux, an imprint of the non-linearity in the
  propagation of the wave. These new terms can be used to build more
  accurate PN templates for the next generation of gravitational wave
  detectors. We give an illustrative estimate of the quantitative
  relevance of the new terms in the orbital phasing of the binary.
 \end{abstract}
 
 \maketitle

\section{Introduction}

The next generation of large interferometric detectors, such as LIGO,
VIRGO, and KAGRA, as well as the future spatial detector eLISA, are
expected to reach the sensitivity required to detect gravitational
waves from the inspiral and coalescence of compact objects binary
systems. Matched filtering techniques used in the data analysis of
these detectors require a very good accuracy of the models built for
the expected signals, which is the main motivation for constructing
higher order post-Newtonian (PN) templates covering the inspiral phase
of the waveform.

The spin of the compact objects, especially of the black holes, has
important quantitative and qualitative effects on the waveforms.
Notably, misaligned spins induce a precession of the binary's
orbital plane. Recent observations indicate that both stellar-size and
supermassive black holes can be generically close to maximally
spinning, and including spin effects in the templates is therefore
relevant.

The spin-orbit, or linear-in-spin contributions, enter the dynamics
and energy flux at the 1.5PN~\footnote{1PN order corresponds to
  $1/c^{2}$, and we will use the notation $\calO(n) =
  \calO(1/c^{n})$. We leave aside quadratic-in-spin contributions,
  which enter at 2PN order.} order for maximal spins. We report here
on new results obtained in a series of
papers~\cite{MBFB13,BMFB13,BMB13,MBBB13}, extending previous work
using the same approach~\cite{FBB06,BBF06}, on the dynamics at the
next-to-next-to-leading, 3.5PN order (finding equivalence with a
previously obtained reduced Hamiltonian in the Arnowitt-Deser-Misner,
or ADM, approach~\cite{HS11so,HSS13}), as well as on the 3.5PN and 4PN order
total energy flux emitted by the binary. These new results can be
directly used for building better PN templates of the inspiral. In the
following, we use the convention $S=cS_{\mathrm{true}}=Gm^{2}\chi$,
with $\chi$ being the dimensionless Kerr parameter, and
  $S_{\mathrm{true}}$ has the dimension of an angular momentum.

\section{Near-zone metric, dynamics and conserved quantities}

The approached used here~\cite{Bliving} is based on the choice of an
harmonic gauge, $\partial_{\nu}h^{\mu\nu} = 0$ with
$h^{\mu\nu}=\sqrt{-g}g^{\mu\nu}-\eta^{\mu\nu}$ being the metric
perturbation, in which Einstein's equations become:
\be\label{eq:Einstein} \Box h^{\mu\nu} = \frac{16 \pi G}{c^{4}} |g|
T^{\mu\nu} + \Lambda^{\mu\nu}(h^{2},h^{3},\dots) \,. 
\ee
The representation of the two compact objects as point particles with spin
is provided, at the linear-in-spin level, by the pole-dipole
model~\cite{Mathisson37repub}. The stress-energy tensor of the
pole-dipole particle model, and the associated equations of motion,
read:
\begin{eqnarray}
	T^{\mu\nu} = c^{2} \int \ud\tau \, p^{(\mu}u^{\nu)}
        \delta(x-z) - c\nabla_{\rho}\int \ud \tau \,
        S^{\rho(\mu}u^{\nu)}\delta(x-z)\,,\label{eq:Tmunu}
        \\ \frac{\uD S^{\mu\nu}}{\ud\tau} = \calO(S^{2}) \,, \quad
        \frac{\uD u^{\mu}}{\ud\tau} =-\frac{1}{2mc}
        R^{\mu}_{\ph{\mu}\nu\rho\sigma} u^{\nu}S^{\rho\sigma} +
        \calO(S^{2}) \label{eq:Papapetrou} \,,
\end{eqnarray}
with $\uD/\ud\tau = u^{\nu}\nabla_{\nu}$. We use the covariant spin
supplementary condition~\cite{Tulczyjew59} $p_{\nu}S^{\mu\nu} =
0$. The mass, defined by $m^{2}=p_{\mu}p^{\mu}$, and the spin norm,
defined by $s^{2} = S_{\mu\nu}S^{\mu\nu}/2$, are conserved.

The iterative solution of~\eqref{eq:Einstein} is parametrized by a set
of metric potentials, $V$, $V_{i}$, $\hat{W}_{ij}$, $\hat{R}_{i}$,
$\hat{X}$, $\hat{Z}_{ij}$, $\hat{Y}_{i}$ and $\hat{T}$, which are
solved for using~\eqref{eq:Tmunu} and lower-order potentials for the
source. As the matter source involves Dirac delta functions,
the gravitational field is singular, and a regularization procedure is
to be specified, for both evaluating the field at the location of the
particles and giving a meaning to divergent integrals. Some of the
metric potentials are computed in the whole near-zone, but others can
be computed only regularized at the location of the two bodies. We
followed the lines of a previous work on the 3PN non-spinning
equations of motion~\cite{BDE04}, and applied the dimensional
regularization (``dimreg'') defined there. We found however that the
``pure Hadamard-Schwartz'' regularization was sufficient at this
order, except for one potential ($\hat{Y}_{i}$) for which we performed
a full computation using dimreg, but we checked that all dimreg
contributions vanish identically in the final result~\cite{MBFB13}.

The equations of evolution of the spins and the equations of motion
are then obtained, at 3PN order and 3.5PN order respectively, by
injecting the regularized metric potentials
in~\eqref{eq:Papapetrou}. With these results in hands, one can look
for a set of conserved quantities, the orbital energy $E$, angular
momentum $J_{i}$, linear momentum $P_{i}$ and center-of-mass integral
$G_{i}$. These results for the dynamics can be checked through the
following important tests:
\begin{itemize}
	\item the existence of a set of conserved quantities $E$,
          $J_{i}$, $P_{i}$, $G_{i}$;
	\item the manifest Lorentz invariance of the obtained equations
          of motion and precession, which must hold in the
          harmonic gauge;
	\item the agreement of the test-mass limit with the known
          equations of motion of a test particle in a Kerr background;
	\item in our case, the existence of a contact transformation
          linking the harmonic-coordinates positions and spins to the
          positions and spins of the ADM approach~\cite{HS11so},
          giving agreement between the two sets of evolution
          equations.
\end{itemize}

The results also include the metric components
themselves~\cite{BMFB13}, either regularized at the particle positions or evaluated at an arbitrary
point of the near-zone, which can be used in other applications. It is more convenient to write the
results in terms of spin vectors with conserved Euclidean norm, which
can be defined in a geometric way~\cite{BMFB13} from the properties
alone that the covariant norm $s$ is conserved and that the spin
tensor is parallel transported.

The equations of motion, of precession as well as the conserved quantities
can be reduced in the center-of-mass frame defined by $G_{i}=0$, and
restricted to the case of quasi-circular orbits of constant radius
except for the effect of the radiation reaction. Defining
$\bm{x}=r\bm{n}$ the orbital separation, $\bm{v}=\dot{\bm{x}}$ the
velocity, $\bm{\ell} = \bm{n}\times\bm{v}/|\bm{n}\times\bm{v}|$ the
normal to the orbital plane, completing the orthonormal triad with
$\bm{\lambda}$, and defining the orbital and precessional frequencies
by $\dot{\bm{n}}=\omega \bm{\lambda}$, $\dot{\bm{\ell}} = -\varpi
\bm{\lambda}$, the structure of the conservative spin-orbit dynamics
is as follows:
\begin{eqnarray}\label{eq:dynamics}
	\ddot{\bm{x}}=\bm{a} &=& - r \omega^{2} \bm{n} + r\omega
        \varpi \bm{\ell}\,, \\ \dot{\bm{S}}_{A} &=&
        \bm{\Omega}_{A}\times \bm{S}_{A} \,, \quad \bm{\Omega}_{A} =
        \Omega_{A}\bm{\ell} \,,
\end{eqnarray}
for $A=1,2$. We refer to the aforementioned articles~\cite{BMFB13} for
the explicit expressions of $\omega$, $\varpi$, $\Omega_{A}$, and we only give here the expression of the additional terms obtained in the
conserved energy. With the further definitions
$\bm{S}=\bm{S}_{1}+\bm{S}_{2}$,
$\bm{\Sigma}=\bm{S}_{2}/m_{2}-\bm{S}_{1}/m_{1}$, $S_{\ell} =
\bm{S}\cdot \bm{\ell}$, $m=m_{1}+m_{2}$, $\nu=m_{1} m_{2}/m^{2}$ and
$\delta m = (m_{1}-m_{2})/m$, in terms of the 1PN parameter
$x\equiv(Gm\omega/c^{3})^{2/3}$, the new 3.5PN spin-orbit
contributions to the energy read:
\begin{eqnarray}\label{eq:energy}
	\delta E_{S} &=&-\frac{m \nu c^2
          x}{2}\left(\frac{x^{3/2}}{G\,m^2}\right) \times
        \nn\\ &&\qquad x^2 \left[
          \left(\frac{135}{4}-\frac{367}{4}\nu+\frac{29}{12}\nu^2\right)S_\ell
          +\left(\frac{27}{4}-39\nu+\frac{5}{4}\nu^2\right)\frac{\delta
            m}{m}\Sigma_\ell\right] +\mathcal{O}\left(8\right)\,.
\end{eqnarray}

\section{Computation of the energy flux}

In a radiative coordinate system, the gravitational waveform can be
written as a multipolar sum over radiative moments
$U_{L},V_{L}$~\cite{Thorne80} which are symmetric and trace-free (STF). The
corresponding energy flux is:
\be\label{flux}
	\mathcal{F} =\sum_{\ell = 2}^{+ \infty} \frac{G}{c^{2\ell +1}}\,\left[
  \frac{(\ell+1)(\ell+2)}{(\ell-1) \ell \, \ell!  (2\ell+1)!!}
  U_L^{(1)} U_L^{(1)} + \frac{4\ell (\ell+2)}{c^2 (\ell-1) (\ell+1)!
    (2\ell+1)!!} V_L^{(1)} V_L^{(1)}\right]\,,
\ee
where $L$ stands for a multi-index $i_{1}\dots i_{\ell}$ and $^{(n)}$ for the
$n^{\mathrm{th}}$ time derivative. In the multipolar post-Newtonian
formalism~\cite{BD86,B98mult}, a systematic iteration of Einstein's
equations in vacuum is first performed, and then the overlap region
between the near-zone and the exterior of the source allows an
asymptotic matching procedure which determines the source and gauge
moments $I_{L},J_{L},W_{L},X_{L},Y_{L},Z_{L}$ parametrizing the
exterior solution, in the form of integrals over the source.

As a result of the matching procedure, the radiative moments contain
contributions of different natures. For instance, we have for the radiative
quadrupole which enters the waveform at the leading order:
\be
	U_{ij} = U_{ij}^{\mathrm{inst}} + U_{ij}^{\mathrm{tails}}
+U_{ij}^{\mathrm{memory}} +U_{ij}^{\text{tails-of-tails}} \,. 
\ee
At leading order, the radiative quadrupole is just the second derivative
of the usual STF mass moment: $U_{ij}^{\mathrm{inst}}=I_{ij}^{(2)} +
\calO(2)$. The non-linearities in the formalism enter through
higher-order instantaneous interactions between multipoles, and by
hereditary contributions such as tails, memory or tail-of-tail
terms. While  the spin-orbit contributions enter the energy
flux at 1.5PN order, our work on the equations of motion and the near-zone
metric allowed a direct computation of the next-to-next-to-leading
contribution, \textit{i.e.} at 3.5PN, where in fact only instantaneous terms arise.

The tails are due to the non-linearity in the propagation of the
outgoing wave. They correspond more formally to the interaction between
the considered multipole and the mass monopole of the system. Taking
the mass quadrupole as an example, the tail contribution is given by
the following hereditary integral extending over the past of the
source:
\be\label{eq:tails} U_{ij}^{\mathrm{tails}}(T_R) = \frac{2 G
  M}{c^3} \int_0^{+\infty}\! \ud\tau \, I_{ij}^{(4)}(T_R-\tau)
\left[\ln \left(\frac{\tau}{2\tau_0} \right)+ \frac{11}{12} \right]\,,
\ee
with $M$ the ADM mass, and $I_{ij}$ the mass quadrupole. The
leading order of the spin-orbit terms due to the tails, at 3PN, was
already studied~\cite{BBF11}.

We addressed the computation of the next-to-leading order, at 4PN. For
dimensional reasons, in the case of quasi-circular orbits, the tails are the only contributions to the flux at 3PN and 4PN
order. The hereditary character of these contributions requires to
control the past dynamics of the binary. It can be shown that
this dynamics can be considered as conservative, but the effect of the
orbital precession needs a priori to be taken into account.

\begin{figure}[t]
	\centering
	\includegraphics[width=0.5\textwidth]{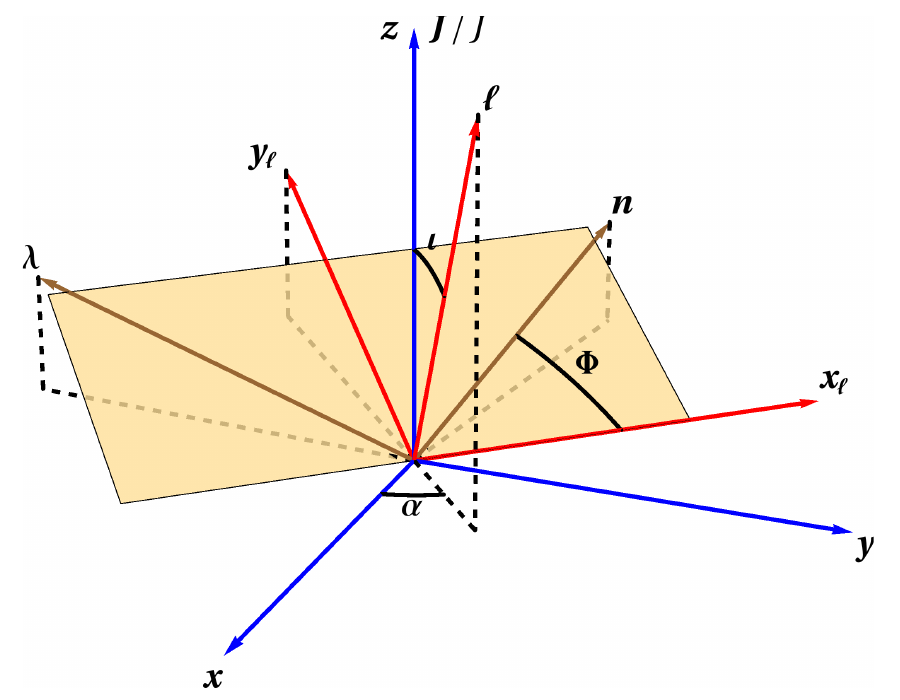}
	\caption{Definition of the Euler angles $\alpha$, $\iota$ and
          $\Phi$. The conserved total angular momentum $\bm{J}$ gives
          a reference direction.}\label{fig:geometry}
\end{figure}

Extending previously obtained results~\cite{BBF11}, we investigated
the precessional effects by formally truncating the dynamics to the
linear-in-spin order, and obtained a solution valid at any PN order when the radiation reaction
is neglected. Defining the Euler angles
$\iota$ and $\alpha$ as in Fig.~\ref{fig:geometry}, we found that the
spin vectors obey (for $A=1,2$):
\be\label{eq:solutionspins}
	\bm{S}_{A}\cdot \bm{\ell} = \cte \,, \quad \bm{S}_{A}\cdot (\bm{n}+\ui
\bm{\lambda}) = S^{A}_{\perp} e^{-\ui \left(  (\omega-\Omega_{A})(t-t_{0}) + \psi^{A}_{0} \right)} \,,
\ee
with $t_{0},S^{A}_{\perp}, \psi^{A}_{0}$ constants, and that precessional
evolution of the moving triad $(\bm{n},\bm{\lambda},\bm{\ell})$ can be
entirely expressed in terms of the quantity:
\be
	\sin \iota \, e^{\ui  \alpha} = - \ui \frac{J_{+}}{|\bm{J}_{\mathrm{NS}}|}e^{\ui \phi} +
\calO(S^{2}) \,,\label{eq:magiccombination}
\ee
where $\phi = \omega(t-t_{0}) + \phi_{0}$ is the orbital phase, and
$J_{+}=\bm{J}\cdot(\bm{n}+\ui \bm{\lambda})$ is linear in spin and has
a simple time dependence obtained from~\eqref{eq:solutionspins}. Thus, the structure of the integrand in~\eqref{eq:tails} is:
\be
	e^{\ui (m \omega + p \Omega_{1} + q \Omega_{2}) t} \,, \quad
\text{with} ~m \in \mathbb{Z} ~\text{and} ~(p,q) \in \{-1,0,1\} \,.
\ee
This simple time dependence allowed a straightforward computation
of the tail contributions to the radiative moments. We found that the
precessional contributions cancel out when combining these moments to
compute the flux, which can be deduced directly from the structure of
the solution for the conservative dynamics, but they do not so in the waveform and
our complete calculation will be useful for a future computation of
the polarizations \footnote{Or, equivalently, the spin-weighted spherical
modes.} of the wave.

The final result for the new (3.5PN+4PN) spin-orbit
contributions to the energy flux reads:
\begin{eqnarray}\label{eq:flux}
	\delta\mathcal{F}_\mathrm{S} \!\!&=&\!\! \frac{32 c^5}{5
          G}\,x^5\,\nu^2\left(\frac{x^{3/2}}{G\,m^2}\right) \times
        \nonumber\\ && \!\!\! \left\{ x^2
        \left[\left(\frac{476645}{6804}+\frac{6172}{189}\nu
          -\frac{2810}{27}\nu^2\right)S_\ell
          +\left(\frac{9535}{336}+\frac{1849}{126}\nu
          -\frac{1501}{36}\nu^2\right)\frac{\delta m}{m}\Sigma_\ell
          \right]\right.\nonumber\\ && \!\!\! \left. + x^{5/2} \left[
          \left( -\frac{3485 \pi}{96} + \frac{13879 \pi}{72}\nu
          \right) S_{\ell} + \left( -\frac{7163 \pi}{672} +
          \frac{130583 \pi}{2016}\nu \right)\frac{\delta m}{m}
          \Sigma_{\ell} \right] +\mathcal{O}\left(6\right) \right\}.
\end{eqnarray}
This result can be checked by testing:
\begin{itemize}
	\item the agreement of the test-mass limit with results form
          the black hole perturbation theory~\cite{TSTS96};
	\item the agreement of the multipoles obtained for one body
          with the ones of a boosted Kerr black hole, computed as
          surface integrals~\cite{BDI04}.
\end{itemize}

\section{The orbital phasing of the binary}

The energy flux can be combined with the result obtained for the
orbital energy~\eqref{eq:energy} through the balance equation
$\mathcal{F} = -\ud E/\ud t$, which can be rewritten as an evolution
equation for the phase and solved using one of the various existing PN
approximants. Table~\ref{tab:cycles} gives the contribution of each
post-Newtonian order to the number of cycles expected to be seen in
ground-based detectors, for typical LIGO/VIRGO targets, using a
``Taylor 2'' approximant. A more complete study by other
authors~\cite{Nitz+13} has investigated the overlaps between templates
built with different approximants, keeping the physical parameters
fixed, including or not these new contributions, and concluded to
their importance.

\begin{table*}[t]
\begin{center}
\caption{Contribution of each post-Newtonian order to the number of
  cycles, for spin-aligned quasi-circular orbits, computed using a
  ``Taylor 2'' approximant between a frequency of $10\mathrm{Hz}$,
  representing the entry in the detector's band, and a cut-off frequency arbitrarily set
  to $x=1/6$. The result is given for typical black hole and neutron-star masses. The parameters $\kappa_{A}$ and $\chi_{A}$ stand for
  the orientation of the spin and the dimensionless Kerr
  parameter. Notice however that the relative importance of these PN
  contributions is affected by the choice of the approximant for the
  phase evolution.\label{tab:cycles}} {\scriptsize
\begin{tabular}{|r|c|c|c|}
\hline LIGO/Virgo & $1.4 M_{\odot} + 1.4 M_{\odot}$ & $10
M_{\odot} + 1.4 M_{\odot}$ & $10 M_{\odot} + 10 M_{\odot}$ \\
\hline Newtonian & $15952.6$ & $3558.9$ & $598.8$ \\ 1PN & $439.5$ &
$212.4$ & $59.1$ \\ 1.5PN & $-210.3+65.6 \kappa_1\chi_1+65.6
\kappa_2\chi_2$ & $-180.9+114.0 \kappa_1\chi_1+11.7 \kappa_2\chi_2$ &
$-51.2+16.0 \kappa_1\chi_1+16.0 \kappa_2\chi_2$ \\ 2PN & $9.9$ & $9.8$
& $4.0$ \\ 2.5PN & $-11.7+9.3 \kappa_1\chi_1+9.3 \kappa_2\chi_2$ &
$-20.0+33.8 \kappa_1\chi_1+2.9 \kappa_2\chi_2$ & $-7.1+5.7
\kappa_1\chi_1+5.7 \kappa_2\chi_2$ \\ 3PN & $2.6-3.2
\kappa_1\chi_1-3.2 \kappa_2\chi_2$ & $2.3 - 13.2\kappa_1\chi_1 - 1.3
\kappa_2\chi_2$ & $2.2-2.6 \kappa_1\chi_1-2.6 \kappa_2\chi_2$ \\ 3.5PN
& $-0.9+1.9 \kappa_1\chi_1+1.9 \kappa_2\chi_2$ & $-1.8+11.1
\kappa_1\chi_1+0.8 \kappa_2\chi_2$ & $-0.8+1.7 \kappa_1\chi_1+1.7
\kappa_2\chi_2$\\ 4PN & $ (\mathrm{NS}) -1.5 \kappa_1\chi_1 - 1.5
\kappa_2\chi_2 $ & $ (\mathrm{NS}) -8.0 \kappa_1\chi_1 - 0.7
\kappa_2\chi_2 $ & $(\mathrm{NS}) -1.5 \kappa_1\chi_1 - 1.5
\kappa_2\chi_2 $\\ \hline
\end{tabular}
}
\end{center}
\end{table*}

\section*{Acknowledgments}

A. Boh\'e is grateful for the support of the Spanish MIMECO grant
FPA2010-16495, the European Union FEDER funds, and the Conselleria
d’Economia i Competitivitat del Govern de les Illes Balears. The last part of this work, on the next-to-leading tail contributions, was realized in collaboration with A. Buonanno from the University of Maryland, College Park, USA.
Our computations were done using Mathematica\textregistered{} and the symbolic tensor calculus package xAct~\cite{xtensor}.

\section*{References}

\bibliographystyle{apsrev4-1.bst}    

\bibliography{/Users/marsat/Documents/publications/bibliographie/ListeRef}

%
%
%
%

\end{document}